\documentclass[prl,twocolumn,superscriptaddress]{revtex4}
\usepackage{amsthm}
\usepackage{amsmath}
\usepackage{revsymb}
\usepackage{latexsym}
\usepackage{amsfonts}
\usepackage{amssymb}
\usepackage{color}
\usepackage{graphicx}
\usepackage{hyperref}
\usepackage{subfigure}
\usepackage{enumitem}
\usepackage{bm}

\newcommand{\id}{\openone} 

\newcommand{\tr}{\mathop{\text{tr}}\nolimits}

\renewcommand{\Re}{\mathop{\text{Re}}\nolimits}

\newcommand{\ket}[1]{|{#1}\rangle}

\definecolor{dgreen}{rgb}{0,0.5,0}

\definecolor{delete}{cmyk}{0.5,0,0,0}

\begin{document}
\title{Remote Parameter Estimation in a Quantum Spin Chain Enhanced by Local Control}
\author{Jukka Kiukas}
\affiliation{Department of Mathematics, Aberystwyth University, Aberystwyth SY23 3BZ, UK}
\author{Kazuya Yuasa}
\affiliation{Department of Physics, Waseda University, Tokyo 169-8555, Japan}
\author{Daniel Burgarth}
\affiliation{Department of Mathematics, Aberystwyth University, Aberystwyth SY23 3BZ, UK}

\date{\today}

\begin{abstract}
We study the interplay of control and parameter estimation on a quantum spin chain. 
A single qubit probe is attached to one end of the chain, while we wish to estimate a parameter on the other end. 
We find that control on the probe qubit can substantially improve the estimation performance and discover some interesting connections to quantum state transfer.
\end{abstract}
\maketitle

\textit{Introduction.---}%
Control and estimation are two sides of the same coin. 
Without control, estimation cannot be performed; without estimation, the parameters required for control are unknown. 
On one side, the statistical inference of parameters specifying partially unknown quantum systems is currently an active research topic \cite{ref:MetrologyScience,ref:HayashiAsymptoticTheory,parisqest,ref:MetrologyNaturePhoto,ref:DowlingReview}. 
In such problems, the control resources for the estimation are usually neglected.
On the other side, in the active field of quantum control \cite{quaint}, the knowledge of the parameters of a given system is often assumed. Given the current world-wide initiatives to build high-performance quantum devices, it is no surprise that the interplay between quantum control and estimation is regaining attention \cite{QSI, Yan, metrologyctrl, QFIctrl}. Only by considering both one has the chance to find an optimal performance. We consider this problem in the context where a large many-body quantum system is observed via a single qubit probe, to estimate an unknown parameter specifying a part of the Hamiltonian localised far away from the probe \cite{carlo,daniel,SoneCappellaro}.
Such a setting is natural in scenarios where a large system is only partially accessible to experimentalists due to limitations forced by the implementation \cite{taminiau}. 
We show that control can substantially improve estimation and reveal some interesting connections with quantum state transfer on spin chains \cite{bose, dbreview}.

\textit{Parameter estimation enhanced by control.---}%
Suppose that we wish to estimate a parameter $\lambda\in\mathbb{R}$ of a large quantum system through a single probe spin in contact with the former.
We are allowed to control and measure the probe, while we have no direct access to the target system.
Our Hamiltonian reads $H_{\mathbf{c},\lambda}(t)=H_\lambda+c(t)H_\text{ctrl}$. The target parameter $\lambda$ is contained in $H_\lambda$, while $H_\text{ctrl}$ represents a local control on the probe, which is tuned by the control pulse $\mathbf{c}=c(t)$. More generally, we could have more control Hamiltonians; however in this paper we only need to consider one. We then let the system evolve from a certain initial state $\ket{\Psi_0}$, e.g., the probe prepared in $\alpha\ket{{\uparrow}}+\beta\ket{{\downarrow}}$ while the target system is initialised in its ground state.
We are given a probing time $T$, during which we control the probe, and we measure the probe in the final state to gain some information on $\lambda$.
We optimise the control field $c(t)$ to enhance the precision of the estimation of $\lambda$.
See Fig.\ \ref{fig:context} for a paradigmatic setup with a spin chain.
\begin{figure}
\includegraphics[width=0.45\textwidth]{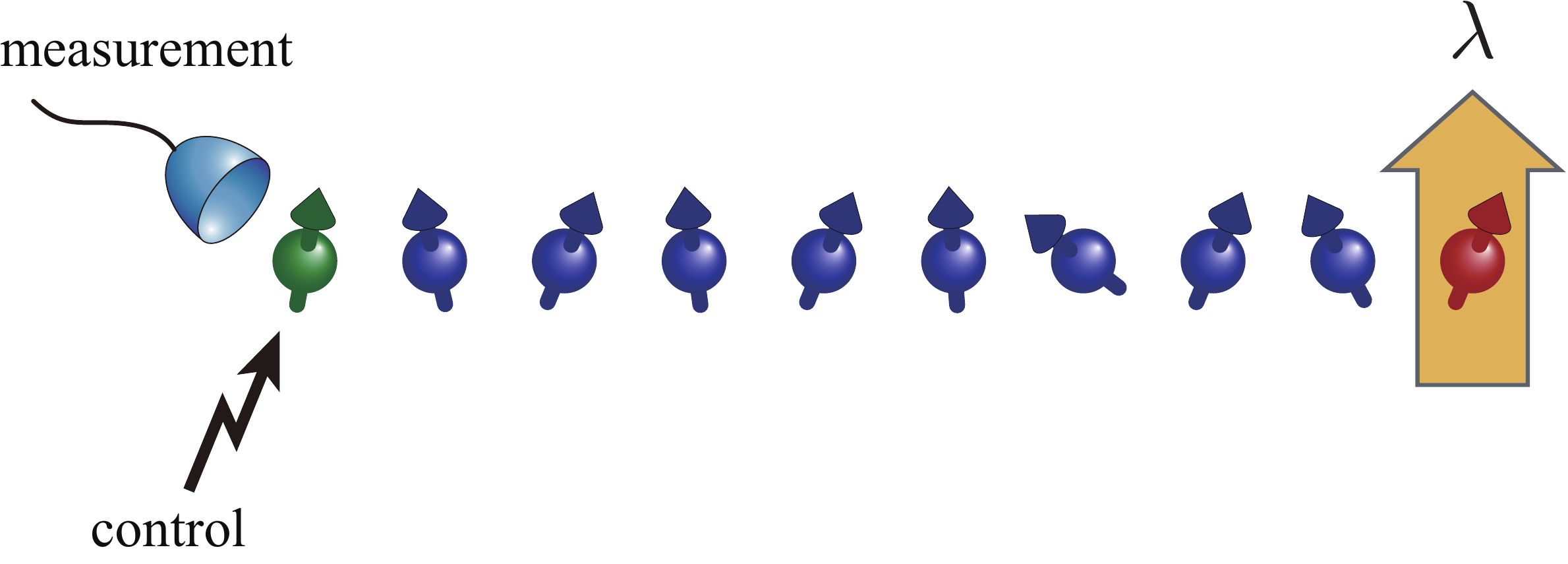}
\caption{(Color online) A large quantum system (e.g., a chain of $N-1$ spins) accessed via a single probe spin. 
The probe can be measured and controlled by switching on and off local Hamiltonians.
We try to enhance the precision of the estimation of an unknown parameter $\lambda$ of the target system (e.g., the strength of a local magnetic field at the end spin) by the control on the probe.
} 
\label{fig:context}
\end{figure}

As we can only measure the probe, the precision of the estimation is ruled by the reduced density operator $\rho_{\mathbf{c},\lambda} = \frac{1}{2} (\id + \mathbf{u}_{\mathbf{c},\lambda}\cdot\bm{\sigma})$ of the probe,
where $\bm{\sigma}=(\sigma_x,\sigma_y,\sigma_z)$ are the Pauli operators and $\mathbf{u}_{\mathbf{c},\lambda}\in\mathbb{R}^3$ is the Bloch vector depending on the target parameter $\lambda$ and the control field $\mathbf{c}$. 
The quantum Fisher information (QFI) \cite{parisqest,ref:MetrologyNaturePhoto} for the estimation of $\lambda$ is then \cite{qubitfisher}
\begin{equation}\label{fisherintext}
F_{\mathbf{c},\lambda}=\|\partial_\lambda \mathbf{u}_{\mathbf{c},\lambda}\|^2+\frac{(\mathbf{u}_{\mathbf{c},\lambda}\cdot\partial_\lambda \mathbf{u}_{\mathbf{c},\lambda})^2}{1-\|\mathbf{u}_{\mathbf{c},\lambda}\|^2}.
\end{equation}
The optimal quantum estimator is given by $E_{\mathbf{c},\lambda} = \lambda \id + L_{\mathbf{c},\lambda} / F_{\mathbf{c},\lambda}$ \cite{parisqest}, where the symmetric logarithmic derivative $L_{\mathbf{c},\lambda}$ has an explicit form given in the Supplemental Material \cite{Suppl}.
It satisfies $\tr[\rho_{\mathbf{c},\lambda} L_{\mathbf{c},\lambda}]=0$ and $F_{\mathbf{c},\lambda} = \tr[\rho_{\mathbf{c},\lambda} L_{\mathbf{c},\lambda}^2]$, showing that the expectation value of $E_{\mathbf{c},\lambda}$ coincides with the target parameter $\lambda$, and the variance $(\Delta E_{\mathbf{c},\lambda})^2=1/F_{\mathbf{c},\lambda}$ saturates the Cram\'er-Rao lower bound on the estimation error  \cite{parisqest,ref:MetrologyNaturePhoto}.
Our basic idea is to tune $\mathbf{c}=c(t)$ to maximise $F_{\mathbf{c},\lambda}$, using a tailored optimal control software.

In practice the estimator $E_{\mathbf{c},\lambda}$ must be constructed with a guessed value of $\lambda$, improved iteratively if necessary. While the general idea of enhancing the QFI by control has appeared recently in different variations (see e.g.\ \cite{Yan,metrologyctrl,QFIctrl}), we have not found an explicit implementation of a feedback-based estimation algorithm. Hence we now proceed to describe a simple one. With the probing time $T$ fixed, the protocol begins with an initial guess $\lambda_0$, followed by the \emph{control step}, where a pulse $\mathbf{c}=\mathbf{c}_\text{max}$ maximising $F_{\mathbf{c},\lambda_0}$ is found. In the next step, $E_{\mathbf{c}_\text{max},\lambda_0}$ is measured on $S_0\sim 1/ (\varepsilon^2 F_{\mathbf{c}_\text{max},\lambda_0} )$ copies of the \textit{true} state $\rho_{\mathbf{c}_\text{max},\lambda_\text{true}}$, and the estimate is updated to $\lambda_{1}=\langle E_{\mathbf{c}_\text{max}, \lambda_0}\rangle$ (the average of the outcomes). The process is then iterated to get successive estimates $\lambda_1,\lambda_2, \lambda_3,\ldots$, and terminated upon reaching a predetermined accuracy $\varepsilon$, in the sense of $|\lambda_{n+1}-\lambda_n|<\varepsilon$. One can also use subsampling to obtain the standard error, which should tend to zero as the estimates $\lambda_n$ converge, providing a better termination condition.
The total number of measurements $S = \sum_n S_n$ needed for the convergence then serves as a measure of the estimation resources; by comparing $S$ with the one obtained by running the protocol without the control step, we can judge the effect of the control on the performance of the estimation.

We use this method, together with simulated measurements (Bernoulli trials) to numerically illustrate the advantage of control in the setting introduced below, where the control turns out to provide a significant reduction of estimation resources for a specific nontrivial model.

\textit{Parameter estimation across a Heisenberg chain.---}%
We consider a 1D chain of $N$ spins with a Hamiltonian
\begin{equation}
H_{\mathbf{c},\lambda}(t) = -\frac{J}{2} \sum_{j=1}^{N-1}
(\bm{\sigma}^{(j)}\cdot\bm{\sigma}^{({j+1})})- c(t)\sigma_z^{(1)}- \lambda\sigma_z^{(N)},
\label{eqn:model}
\end{equation}
where $\bm{\sigma}^{(j)}=(\sigma_x^{(j)},\sigma_y^{(j)},\sigma_z^{(j)})$ are the Pauli matrices for the $j$th spin, $J\,(>0)$ is a fixed coupling constant, $c(t)$ is a control field on the first (probe) spin (magnetic field at the first site), and $\lambda$ is the unknown target parameter (magnetic field at the last site); 
see Fig.\ \ref{fig:context}. This setting has been shown to provide full controllability over the whole chain \cite{fullcont}, and its state transport properties have been studied in \cite{bose} without control. For quantum estimation in spin systems see e.g.\ \cite{daniel, loschmidt, fisherspinchain}.

For each spin, the eigenstates of $\sigma_z$ are denoted by $\ket{{\uparrow}}$ and $\ket{{\downarrow}}$. The first spin ($j=1$) is the probe, and we initialise it in $a\ket{{\uparrow}} + b\ket{{\downarrow}}$, with all the other spins in $\ket{{\downarrow}}$.  
Since $H_{\mathbf{c},\lambda}(t)$ commutes with the $z$ component of the total spin $\sum_{j=1}^N\sigma^{(j)}_z$, the system remains in the subspace spanned by $\ket{0} := \ket{{\downarrow}}\otimes\cdots \otimes \ket{{\downarrow}}$, where all the spins are in $\ket{{\downarrow}}$, and $\ket{j} := \ket{{\downarrow}}\otimes\cdots \otimes \ket{{\downarrow}}\otimes\ket{{\uparrow}}\otimes \ket{{\downarrow}} \otimes \cdots \otimes \ket{{\downarrow}}$, where only the $j$th spin ($j=1,\ldots,N$) is flipped to $\ket{{\uparrow}}$. The initial state of the chain is therefore $\ket{\Psi_0}=a\ket{0} + b\ket{1}$.
We fix a total probing time $T$, and discretise it into $m$ time slots, each of which having a different constant control. Hence the pulse $\mathbf{c}=c(t)$ is given by the vector $\mathbf{c}=(c_1,\ldots, c_m)\in \mathbb R^m$ determining the final state $\ket{\Psi_{\mathbf{c},\lambda}(T)} = e^{-iH_{c_m,\lambda}T/m}\cdots e^{-iH_{c_1,\lambda}T/m}\ket{\Psi_0}$. We reduce this total state to the state $\rho_{\mathbf{c},\lambda}$ of the probe spin, to which our general estimation procedure can be applied.

\textit{Analytical study for the two-spin case.---}%
Before going to numerics, let us first look at the two-spin case ($N=2$) to understand how the control helps the estimation; this is already a nontrivial indirect estimation problem.

By recalling the Trotter formula, the unitary transformation induced by $H_{\mathbf{c},\lambda}(t)$ in (\ref{eqn:model}) for a generic $N$ is basically composed of the single-spin rotations on the last spin by $e^{i\lambda\sigma_z^{(N)}t}$ with the unitary transformations by the other parts of $H_{\mathbf{c},\lambda}(t)$ inserted in between.
As discussed in \cite{vittorio}, the QFI for the estimation of $\lambda$ embedded by such sequential transformations is upper bounded by the optimal QFI for the estimation of $\lambda$ embedded solely by $e^{i\lambda\sigma_z^{(N)}T}$ with direct initialisation and measurement on the last spin permitted.
That is, the QFI for our problem is upper bounded by $F_{\mathbf{c},\lambda}\le4T^2$.

If we are allowed to access all the spins, initialising and measuring them at will, this upper bound is actually achievable, by initialising and measuring the so-called NOON state $(\ket{{\downarrow}}\otimes \cdots \otimes\ket{{\downarrow}}+\ket{{\uparrow}}\otimes \cdots \otimes\ket{{\uparrow}})/\sqrt{2}$, with no control $c(t)$ required during the probing.
In our problem, however, only the probe spin is accessible, and it is not clear whether this upper bound is reachable.
In particular, this puts a strong constraint on the maximal achievable QFI in the uncontrolled case. 
For the two-spin case, it is easy to compute the QFI in (\ref{fisherintext}) for the probe spin in the absence of the control, and to perform the optimisation of $a$ and $b$ in the initial state $\ket{\Psi_0}$. We find the asymptotic scaling of the optimised QFI for large $T$ given by $T^2/[(1-\lambda^2/J^2)(1+\lambda^2/J^2)^2]$ for $\lambda^2<1/2$ and $4T^2(\lambda^2/J^2)/(1+\lambda^2/J^2)^2$ otherwise. This is the best we can attain for the two-spin case without control. This QFI is upper bounded by $T^2$, and is smaller than the above upper bound $4T^2$ at least by a factor of $4$.

The control on the probe can improve the estimation; we can even get close to the upper bound $4T^2$ if $N$ is not large.
Indeed, we provide a naive protocol consisting of three steps, and show that it is asymptotically optimal for the two-spin case.
The first step is to use the control to remotely prepare a good state for sensing. 
The second step is to use the control to let the system evolve so as to acquire the parameter as much as possible.  
The final step is to remotely measure the state, by mapping the state into one that is measurable locally at the probe.

In the first step, we start with the initial state $\ket{\Psi_0}=\ket{1}=\ket{{\uparrow}{\downarrow}}$, and keep the control field at $c(t)=\lambda$ for time $\pi/(4J)$. 
This prepares $(\ket{{\uparrow}{\downarrow}}+i\ket{{\downarrow}{\uparrow}})/\sqrt{2}$ up to a global phase.
In the second step, we apply a very strong field $c(t)\gg J$ to suppress the exchange interaction between the two spins for time $t=T-\pi/(2J)$ \cite{benjamin}. 
Up to a global phase, the system evolves into $(\ket{{\uparrow}{\downarrow}}+ie^{2i\lambda t}\ket{{\downarrow}{\uparrow}})/\sqrt{2}$, acquiring the relative phase depending on the target parameter $\lambda$. 
In the final step, we apply a strong pulse to induce an instantaneous rotation on the probe spin around the $z$ axis to cancel the relative phase $e^{2i\lambda t}$, let the system evolve with $c(t)=\lambda$ for time $\pi/(4J)$, and measure $\sigma_z^{(1)}$ of the probe.
This effectively amounts to measuring $(\ket{{\uparrow}{\downarrow}}\pm ie^{2i\lambda t}\ket{{\downarrow}{\uparrow}})/\sqrt{2}$ in the state after the second step, which is an optimal measurement to estimate the relative phase.
Note that the final state just before the measurement is $\ket{{\downarrow}{\uparrow}}$.
This does not appear to depend on $\lambda$, but the value of $\lambda$ taken in the first and last steps is just a guess and can be different from the true value.
If the guess is not perfect, the probe fails to become the pure state $\ket{{\downarrow}}$.
This failure can be detected by the measurement of $\sigma_z^{(1)}$, which helps us learn about the parameter $\lambda$.
The QFI by the above procedure is calculated to be $F_{\mathbf{c},\lambda}=4[T-(\pi/2-1)/J]^2$.
Since the time $\pi/(2J)$ spent for the first and last steps is finite, it becomes negligible for large $T$, and the QFI asymptotically approaches the ultimate bound $F_{\mathbf{c},\lambda}\to4T^2$.
In this way, the control can enhance the estimation significantly.

\textit{Numerical simulation for a longer chain.---}%
Let us now look at longer chains. We implement the crucial control step of the above estimation procedure using the optimal control software \textit{QTRL}, which is a part of the QuTip control package \cite{qutip,qtrl}. 
In addition to the existing software, we have implemented an exact gradient of the QFI in a way applicable to an arbitrary quantum system probed by a single spin.
See the Supplemental Material \cite{Suppl} for details. 
We note that this requires the computation of the second derivatives of matrix exponentials, for which we have employed the method given in \cite{gradctrl}.

In Fig.\ \ref{fig:gain}, the QFI $F_{\mathbf{c},\lambda}$ normalised by $T^2$ with optimised control is shown as a function of the probing time $T$ for the chain of $N=5$ spins, and is compared with the uncontrolled case. The initial state is $\ket{\Psi_0}=\ket{1}$ for the controlled case, while it is optimised to maximise the QFI for each $T$ for the uncontrolled case.
It is clear from this result that the local control on the probe improve the precision of the estimation. In order to illustrate this explicitly, we display in the inset the results of the full estimation algorithm (described above) for the specific probing time $T=13.5/J$: we observe that the estimation resource, i.e., the number of measurements $S$ needed for the convergence, is significantly reduced by control, even when taking into account the considerable random variation in the results.

\begin{figure}
\begin{tabular}{ll}
\footnotesize &\\[-3.3truemm]
&\includegraphics[width=0.43\textwidth]{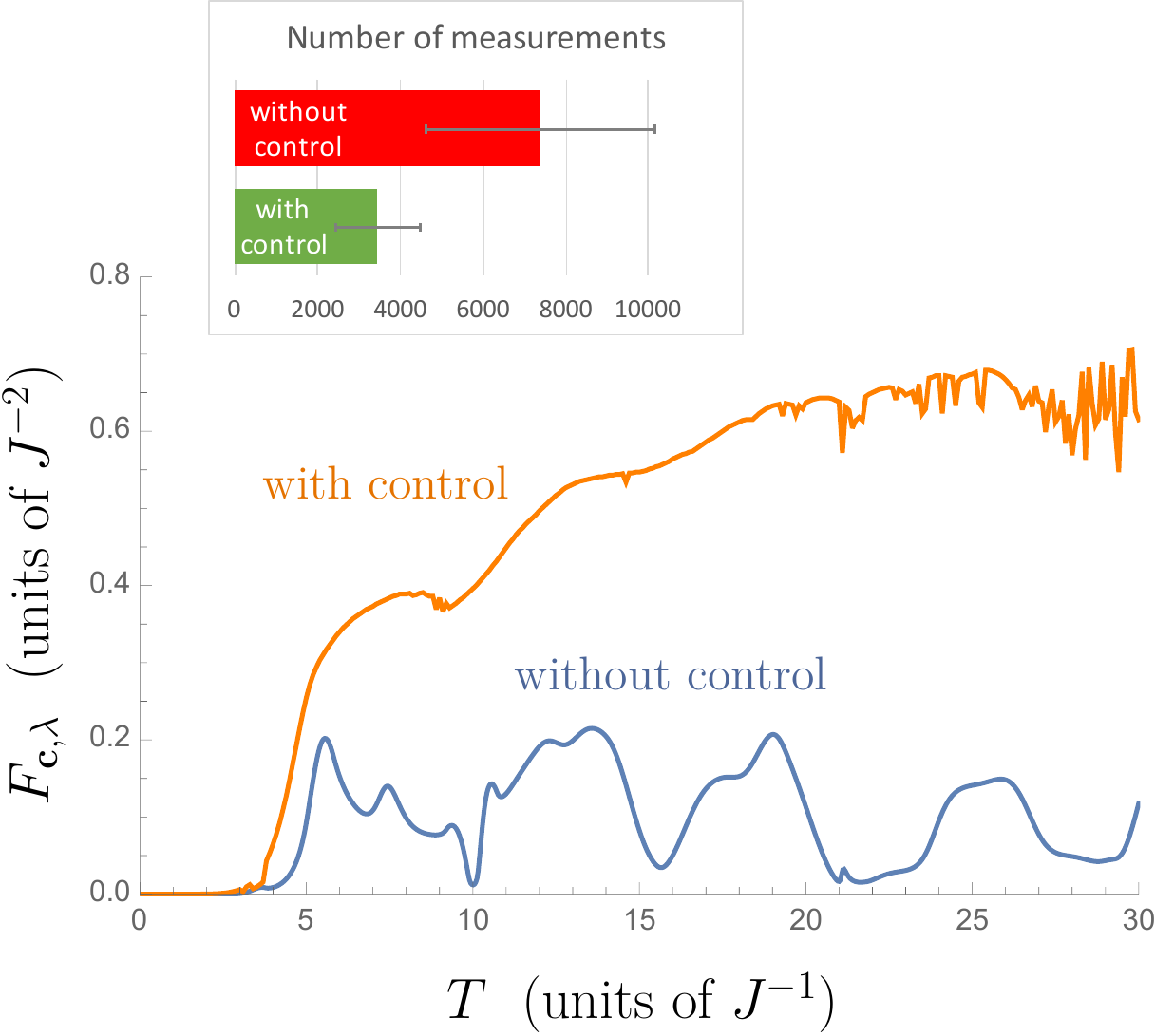}\\
\end{tabular}
\caption{(Color online) 
Estimation of the target value $\lambda_\text{true}=0$ for $N=5$ spins.
The QFI with optimised control is normalised by $T^2$ and shown as a function of the probing time $T$ (orange).
The initial state is chosen to be $\ket{\Psi_0}=\ket{1}$.
Each probing time $T$ is divided into 20 time slots for the control pulses $\mathbf{c}$ to be optimised.
For each $T$, 20 numerical experiments are performed with randomly chosen initial control pulses $\mathbf{c}_0$, and the optimal QFI is taken among them.
The QFI normalised by $T^2$ achievable without control is also shown for comparison (blue).
For this uncontrolled case, the initial state $\ket{\Psi_0} = a|0\rangle+b|1\rangle$ is optimised for each $T$.
Inset: The total number of measurements $S$ needed to reach the target value $\lambda_\text{true}=0$ with accuracy $\varepsilon=0.01J$ starting from initial guess $\lambda_0 = 0.1J$, with and without optimal control. The probing time is $T=13.5/J$.
The value is the average over $500$ numerical runs with the associated standard deviation indicated by an error bar.
}
\label{fig:gain}
\end{figure}

Longer chains naturally require longer probing times $T$, as ruled by the fundamental Lieb-Robinson bound for the propagation speed of the excitation, which for this specific system has been studied in \cite{bose}, and found to increase linearly in $N$. From the estimation point of view, the relevant threshold $T_0$ is the time after which we start gaining the information on $\lambda$; for the $N=5$ case, we observe from Fig.\ \ref{fig:gain} that $T_0\sim4/J$. Another interesting time would be the one after which the Fisher information rate $F_{\mathbf{c},\lambda}/T^2$ no longer increases essentially, and the relevant question is how close to the ultimate precision bound $4$ we can get. While in the two-spin case we could construct a control sequence asymptotically achieving the bound, it is now more difficult to get close to it. Loosely speaking this is because transferring the excitation to the end of the chain takes a longer time, and keeping it there to acquire the information is more difficult as the control is at the other end of the chain. Furthermore, as we see from Fig.\ \ref{fig:gain}, the numerics appears to become unstable for large values of $T$, in that different initial pulses $\mathbf{c}_0$ lead to different rates, making it difficult to judge if we have actually found the maximal rate. For the $N=5$ case, we observe from Fig.\ \ref{fig:gain} that after $T=20/J$, the rate is roughly $F_{\mathbf{c},\lambda}/T^2\sim 0.6$. With $N=10, 15, 20$ we found the best rates $F_{\mathbf{c},\lambda}/T^2\sim 0.1, 0.05, 0.02$, respectively ($F_{0,\lambda}/T^2\sim 0.07, 0.03, 0.01$, respectively, in the uncontrolled case). While the QFI drops quickly with increasing the distance between the probe and the target field, the control increases it significantly.

\begin{figure}
\begin{tabular}{ll}
\footnotesize(a)&\\[-3.3truemm]
&\includegraphics[width=0.44\textwidth]{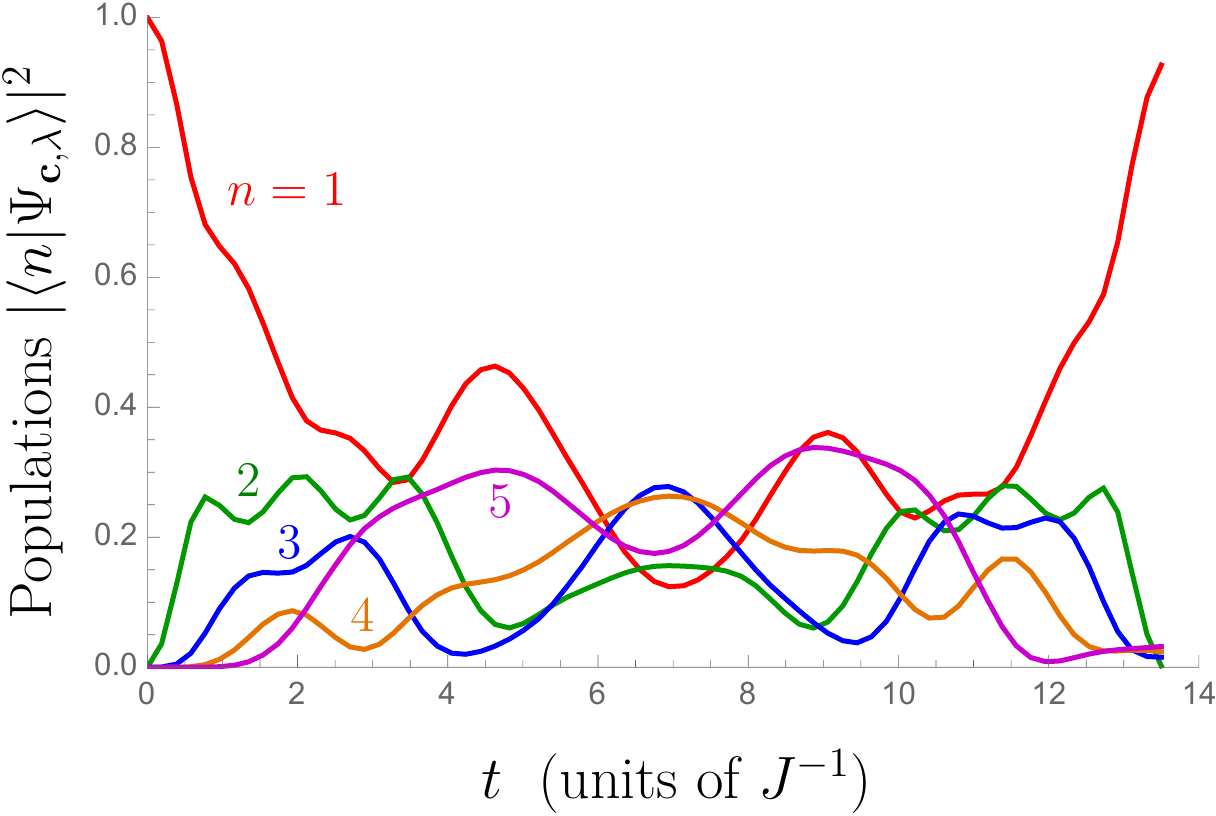}\\
\footnotesize(b)&\\[-3.3truemm]
&\includegraphics[width=0.44\textwidth]{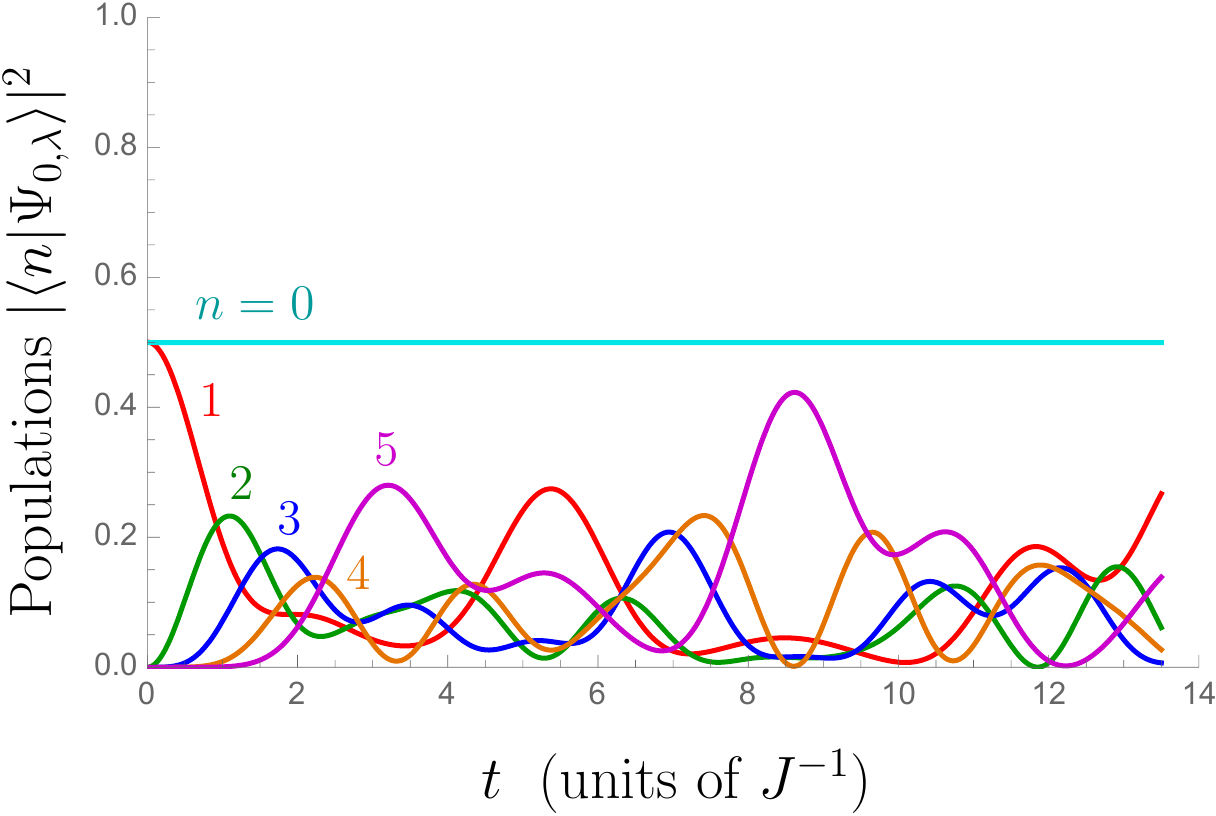}
\end{tabular}
\caption{(Color online) 
Evolution of the population at each site of the chain with $N=5$ spins for the estimation of $\lambda_\text{true}=0$, (a) with an optimal control and (b) with no control, for probing time $T=13.5/J$ divided into $m=70$ time slots.
The initial state is chosen to be $\ket{\Psi_0}=\ket{1}$ for the controlled case (a), while it is optimised to $\ket{\Psi_0}=\cos(\theta/2)\ket{0}+e^{i\varphi}\sin(\theta/2)\ket{1}$ with $(\theta,\varphi)=(1.57153, 2.41026)$ for the uncontrolled case (b).
This clearly shows the structured character in the controlled case, in contrast to the uncontrolled case.
In the controlled case, the excitation propagates across the chain and returns to the first site, facilitating the information transfer.} 
\label{fig:exdynamics}
\end{figure}

Let us then see how the excitation is transferred across the chain.
In Fig.\ \ref{fig:exdynamics}, the evolution of the population probability at each site is shown for the probing time $T=13.5/J$. In the controlled case, we start with $\ket{\Psi_0}=\ket{1}$. As in the two-spin case, we observe roughtly three stages: first the excitation propagates across the chain, reaching the end at around $t\sim2/J$. While the information on $\lambda$ is accumulated, the populations remain more or less the same with some fluctuations, and eventually the excitation is brought back to the probe spin. We observed that this behaviour is fairly typical also with other parameters and longer chains. In particular, the system often returns to the initial state $\ket{1}$ by the time the measurement is performed on the probe spin. As we already discussed in the two-spin case, the final state $\ket{1}$ does not explicitly depend on the target parameter $\lambda$; we learn about $\lambda$ by the failure of the return of the excitation.

From the numerics it appears that the initial state $|1\rangle$ gives the best estimation results. In order to understand this, let us exclude a couple of suggestive alternatives: one might think that a superposition state of the probe such as $(\ket{0}+e^{i\phi}\ket{1})/\sqrt{2}$ would be good for sensing the phase $\phi$. Such a final pure state, as a reduced state of the probe spin, is not available for the initial state $\ket{1}$, since the probe spin is entangled with the rest of the chain when the probe spin is partially populated. It is also understandable that initial states $\ket{\Psi_0}=a\ket{0}+b\ket{1}$ with small $b$ are not useful, since the excitation transferred to the target spin is small.

\textit{Discussion and outlook.---}%
We started with a natural setup for controlled parameter estimation in a quantum many-body system and found that its optimal performance is closely related to achieving perfect state transfer \cite{bose}.
Although the methods developed here were used for a specific system, they are completely general. This means that we can immediately employ the program to study spin networks and experimental setups, paving the way to an optimal interplay of control and estimation in quantum technologies.

\begin{acknowledgments}
We acknowledge fruitful discussions with Rafa{\l}  Demkowicz-Dobrza{\'n}ski and Ugo Marzolino. 
This work was supported by the Top Global University Project from the Ministry of Education, Culture, Sports, Science and Technology (MEXT), Japan.
DB acknowledges support from the EPSRC Grant No.\ EP/M01634X/1.
KY was supported by the Grant-in-Aid for Scientific Research (C) (No.\ 26400406) from the Japan Society for the Promotion of Science (JSPS) and by the Waseda University Grant for Special Research Projects (No.\ 2016K-215).
\end{acknowledgments}


\setcounter{equation}{0}
\newpage
\mbox{}
\newpage
\title{Supplemental Material:\\Remote Parameter Estimation in a Quantum Spin Chain Enhanced by Local Control}
\author{Jukka Kiukas}
\affiliation{Department of Mathematics, Aberystwyth University, Aberystwyth SY23 3BZ, UK}
\author{Kazuya Yuasa}
\affiliation{Department of Physics, Waseda University, Tokyo 169-8555, Japan}
\author{Daniel Burgarth}
\affiliation{Department of Mathematics, Aberystwyth University, Aberystwyth SY23 3BZ, UK}
\maketitle

\section{Supplemental Material}
\subsection{Optimal Estimation of a Single Qubit}
After tracing out the large system as described in the main text, we are left with the probe qubit with the density matrix
\[
\rho_{\mathbf{c},\lambda} = \frac 12 (\id + \mathbf{u}_{\mathbf{c},\lambda}\cdot\bm{\sigma}),
\]
where $\lambda\in \mathbb R$ is the \emph{unknown parameter} to be estimated, and $\mathbf c = c(t)$ is a control pulse to be optimised for optimal QFI\@. For any parameter value $\lambda$, the optimal measurement is given by 
\[
E_{\mathbf{c},\lambda} = \lambda \id + L_{\mathbf{c},\lambda} / F_{\mathbf{c},\lambda},
\]
where the symmetric logarithmic derivative $L_{\mathbf{c},\lambda}=\alpha_{\mathbf{c},\lambda}\id +\mathbf{v}_{\mathbf{c},\lambda}\cdot \bm{\sigma}$ is determined by the coefficients
\begin{align*}
\alpha_{\mathbf{c},\lambda} & = -\frac{\mathbf{u}_{\mathbf{c},\lambda}\cdot \partial_\lambda\mathbf{u}_{\mathbf{c},\lambda}}{1-\|\mathbf{u}_{\mathbf{c},\lambda}\|^2},\\
\mathbf{v}_{\mathbf{c},\lambda}
&= \partial_\lambda \mathbf{u}_{\mathbf{c},\lambda}+\frac{\mathbf{u}_{\mathbf{c},\lambda}\cdot \partial_\lambda \mathbf{u}_{\mathbf{c},\lambda}}{1-\|\mathbf{u}_{\mathbf{c},\lambda}\|^2} \mathbf{u}_{\mathbf{c},\lambda}.
\end{align*}
It is easy to verify that the Hermitian operator $L_{\mathbf{c},\lambda}$ admits a projective measurement with two outcomes 
\[
\ell^{\pm}_{\mathbf c,\lambda}= \alpha_{\mathbf{c},\lambda} \pm \|\mathbf{v}_{\mathbf{c},\lambda}\|
\]
and projectors $(\id \pm \bm{\sigma}\cdot \mathbf{v}_{\mathbf{c},\lambda}/\|\mathbf{v}_{\mathbf{c},\lambda}\|)/2$.

In the estimation procedure described in the main text, the measurement of $E_{\mathbf{c}_\text{max},\lambda_n}$ with an optimal control pulse $\mathbf{c}_\text{max}$ for the $n$th \emph{guessed value} $\lambda_n$ is performed on the state $\rho_{\mathbf{c}_\text{max},\lambda_\text{true}}$, which is the \emph{true} quantum state generated with that pulse. Hence the probabilities for the two outcomes of the measurement of $E_{\mathbf{c}_\text{max},\lambda_n}$ are given by
\[
p_{\pm} = \frac{1}{2}(1\pm \mathbf{u}_{\mathbf c_\text{max}, \lambda_\text{true}}\cdot\mathbf{v}_{\mathbf{c}_\text{max},\lambda_n}/\|\mathbf{v}_{\mathbf{c}_\text{max},\lambda_n}\|).
\]
By repeating the measurement on $S_n\sim 1/ (\varepsilon^2 F_{\mathbf{c}_\text{max},\lambda_n})$ copies of the true state, we would estimate the expectation value
\begin{equation}\label{expe}
\langle E_{\mathbf{c}_\text{max}, \lambda_n}\rangle = \lambda_n+ (\ell^+_{\mathbf{c}_\text{max}, \lambda_n} p_++\ell^-_{\mathbf{c}_\text{max}, \lambda_n} p_-)/F_{\mathbf c_\text{max}, \lambda_n}.
\end{equation}
We simulate this experiment by generating random outcomes from a Binomial distribution with sample size $S_n$ and the true (unknown!) probabilities $p_\pm$ calculated from the above formula. This gives us frequencies $\tilde p_{\pm}$ which we then use in place of the true probabilities $p_{\pm}$ in the above formula to calculate updated estimate $\lambda_{n+1}=\langle E_{\mathbf{c}_\text{max}, \lambda_n}\rangle$ described in the main text.

\subsection{Computation of the QFI and Its Gradient}
The optimal control software \emph{QTRL} \cite{qtrl} generally optimises control pulses by minimising a given fidelity, which by default is chosen to be a suitable distance from a target quantum object (either a state or a unitary operator). For the purpose of the present work, we needed to implement a completely different fidelity, namely the negative of the QFI $F_{\mathbf{c},\lambda}$ described in the main text. Since the optimisation is based on a gradient search, the implementation requires the computation of both the function
\[
\mathbf{c} \mapsto F_{ \mathbf{c}, \lambda}
\]
and its gradient. 
We now proceed to detail how these are obtained. We note that even though we applied the scheme to the specific Heisenberg spin chain, the composed optimisation program is general, and can in principle be used in any setting involving a large system estimated via an embedded single qubit probe.

First of all, we need to specify how the probe is embedded. This is done simply by defining how the qubit Pauli matrices act on the total system consisting of both the probe and the large system. For instance, if the qubit is a subsystem (i.e., a tensor factor), the action of the $i$th Pauli matrix is just $\tilde\sigma_i = \sigma_i \otimes \id$, where the nontrivial action is on the probe system and the identity acts everywhere else. In our case, working in a single excitation sector, it is necessary to consider direct sum instead; hence we allow arbitrary embeddings $\sigma_i \mapsto \tilde\sigma_i$.

As described in the main text, the final state is determined by a control pulse $\mathbf{c}=(c_1,\ldots,c_m)\in \mathbb R^m$, through $\ket{\Psi_{\mathbf{c},\lambda}(T)} = U_{\lambda, \mathbf{c}}\ket{\Psi_0}$, where the total transformation is a unitary $U_{\mathbf{c},\lambda} = U_{c_m,\lambda} \cdots U_{c_1,\lambda}$ given by the product of unitary propagators of the form $U_{c,\lambda} = e^{-iH_{c,\lambda}T/m}$. Here the generator $H_{c,\lambda}=H_\lambda + c H_\text{ctrl}$ depends on the target parameter $\lambda$ and a single constant $c$ characterising the strength of the applied control. By tracing out everything except the probe qubit, we obtain the Bloch vector $\mathbf{u}_{\mathbf{c}, \lambda}$ of the probe, which gives us the QFI $F_{\mathbf{c},\lambda}$ via Eq.\ (1) of the main text. The control optimisation refers to maximising $F_{\mathbf{c},\lambda}$ with respect to the pulse $\mathbf{c}$. More generally, the program can handle generators of the form $H_{c^{(1)}, \ldots, c^{(M)}, \lambda} = H_\lambda + \sum_{k=1}^M c^{(k)} H^{(k)}_\text{ctrl}$, i.e., with several control pulses, but we present the following with a single pulse so as to avoid cluttering the notation.

In order to compute $F_{\mathbf{c},\lambda}$, we need the components of the Bloch vector $\mathbf{u}_{\mathbf{c}, \lambda}$ of the final probe state, together with their derivatives:
\begin{align*}
\mathbf u_{\mathbf{c},\lambda}  &= \langle U_{\mathbf{c},\lambda}\Psi_0| \tilde{\bm{\sigma}} | U_{\mathbf{c},\lambda}\Psi_0\rangle,\\
\partial_\lambda \mathbf{u}_{\mathbf{c},\lambda} &= 2 \Re\langle \partial_\lambda U_{\mathbf{c},\lambda}\Psi_0|\tilde{\bm{\sigma}} |U_{\mathbf{c},\lambda}\Psi_0\rangle,
\end{align*}
where $\tilde{\bm{\sigma}}=(\tilde \sigma_1,\tilde \sigma_2,\tilde \sigma_3)$ and the expectation values on the right-hand side are understood componentwise in the obvious fashion. Using the chain rule we can write the relevant derivative as
\[
\partial_\lambda U_{\mathbf{c},\lambda}=\sum_{i=1}^{m-1} U_{c_m,\lambda}\cdots U_{c_{i+1},\lambda} (\partial_\lambda U_{c_i,\lambda}) U_{c_{i-1},\lambda}\cdots U_{c_1,\lambda},
\]
which we get once we have a method of computing $U_{c,\lambda}$ and $\partial_\lambda U_{c,\lambda}$ for arbitrary $c$.

We now proceed to look at the gradient. We use the shorthand notation $\partial_{i}=\frac{\partial}{\partial c_{i}}$ and suppress the parameter dependence for simplicity. By differentiating the formula (1) in the main text, we find
\begin{align*}
\partial_{i} F_{\mathbf{c},\lambda} ={}&2(\partial_{i}\partial_\lambda \mathbf{u})\cdot\partial_\lambda \mathbf{u}\\
&{}+ 2g(\partial_{i}\partial_\lambda\mathbf{u}\cdot \mathbf{u}+ \partial_\lambda \mathbf{u}\cdot \partial_{i}\mathbf{u} + g\,\partial_{i}\mathbf{u}\cdot \mathbf{u}),
\end{align*}
where $g=(\partial_\lambda \mathbf{u}\cdot \mathbf{u})/(1-\|\mathbf{u}\|^2)$. In order to evaluate this function, we need two new quantities
\begin{align*}
\partial_i \mathbf u ={}& 2 \Re\langle \partial_i U_{\mathbf{c},\lambda}\Psi_0|\tilde{\bm{\sigma}} |U_{\mathbf{c},\lambda}\Psi_0\rangle,\\
\partial_{i}\partial_\lambda \mathbf u ={}& 2 \Re\langle \partial_i \partial_\lambda U_{\mathbf{c},\lambda}\Psi_0|\tilde{\bm{\sigma}} |U_{\mathbf{c},\lambda}\Psi_0\rangle\\
&{}+ 2\Re\langle \partial_\lambda U_{\mathbf{c},\lambda}\Psi_0|\tilde{\bm{\sigma}} |\partial_i U_{\mathbf{c},\lambda}\Psi_0\rangle.
\end{align*}
Since $c_i$ only appears in the $i$th propagator (while $\lambda$ is in each one), we have
\begin{align*}
&\partial_{i} U_{\mathbf{c},\lambda} = U_{c_1,\lambda}\cdots U_{c_{i-1},\lambda} (\partial_{i} U_{c_i,\lambda})U_{c_{i+1},\lambda}\cdots U_{c_m,\lambda},\\
&\partial_{i} \partial_\lambda U_{\mathbf{c},\lambda}
\nonumber\\
&\quad
= (\partial_\lambda
U_{c_1,\lambda}) \cdots U_{c_{i-1},\lambda}(\partial_{i} U_{c_i,\lambda}) U_{c_{i+1},\lambda}\cdots U_{c_m,\lambda}\\
&\qquad
{}+ \cdots
+U_{c_1,\lambda}\cdots U_{c_{i-1},\lambda}(\partial_{i} \partial_\lambda U_{c_i,\lambda}) U_{c_{i+1},\lambda}\cdots U_{c_m,\lambda}\\
&\qquad
 {}+\cdots
+U_{c_1,\lambda}\cdots U_{c_{i-1},\lambda}(\partial_{i} U_{c_i,\lambda})U_{c_{i+1},\lambda}\cdots\partial_\lambda U_{c_m,\lambda}.
\end{align*}
Hence, in order to compute both $F_{c,\lambda}$ and its gradient, we need the four quantities $U_{c,\lambda}$, $\partial_{i} U_{c,\lambda}$, $\partial_\lambda U_{c,\lambda}$, and $\partial_i\partial_\lambda U_{c,\lambda}$. While the first two had already been implemented in \emph{QTRL}, we needed to construct the $\lambda$-derivatives specific to the estimation context.

Following \cite{gradctrl_suppl}, we observe that all these derivatives can be conveniently computed simultaneously as follows. We write $W_{t} = e^{-i H_{c,\lambda}t}$ for each $t$, and successively differentiate the evolution equation to get
\begin{align*}
i\partial_t W_{t} &= H_{c,\lambda}W_{t}\\
i\partial_t \partial_\lambda W_{t} &= (\partial_\lambda H_\lambda)W_{t} + H_{c,\lambda}\partial_\lambda W_{t}\\
i\partial_t \partial_c W_{t} &= H_\text{ctrl}W_{t} + H_{c,\lambda}\partial_c W_{t}\\
i\partial_t \partial_c\partial_\lambda W_{t} &= H_\text{ctrl} \partial_\lambda W_{t} + (\partial_\lambda H_\lambda)\partial_c W_{t}+ H_{c,\lambda} \,\partial_c\partial_\lambda W_{t}.
\end{align*}
This can be written as the single block matrix equation
\[
i \partial_t
\begin{pmatrix}W_{t}\\
\partial_\lambda W_{t}\\
\partial_c W_{t}\\
\partial_{c}\partial_\lambda W_{t}
\end{pmatrix}
= \mathbf M_{c,\lambda}
\begin{pmatrix}W_{t}\\
\partial_\lambda W_{t}\\
\partial_c W_{t}\\
\partial_{c}\partial_\lambda W_{t}
\end{pmatrix}
\]
with
\[
\begin{pmatrix}W_{0}\\
\partial_\lambda W_{0}\\
\partial_c W_{0}\\
\partial_{c}\partial_\lambda W_{0} 
\end{pmatrix} = \begin{pmatrix}\id\\ 0\\0\\0
\end{pmatrix},
\]
where
$$
\mathbf M_{c,\lambda}=\begin{pmatrix}
H_{c,\lambda} & 0 & 0 & 0\\
\partial_\lambda H_\lambda & H_{c,\lambda} & 0 & 0\\
H_\text{ctrl} & 0 & H_{c,\lambda} & 0\\
0 & H_\text{ctrl} & \partial_\lambda H_\lambda & H_{c,\lambda}
\end{pmatrix}.
$$
Since $U_{c,\lambda}= W_{T/m}$, we immediately get
$$
\begin{pmatrix}U_{c,\lambda}\\
\partial_\lambda U_{c,\lambda}\\
\partial_c U_{c,\lambda}\\
\partial_{c}\partial_\lambda U_{c,\lambda}
\end{pmatrix} = e^{-i\mathbf M_{c,\lambda}T/m} \begin{pmatrix}\id\\ 0\\0\\0
\end{pmatrix},
$$
that is, we can extract all the required derivatives from the first column of the blocks of the matrix
$e^{-i\mathbf M_{c,\lambda}T/m}$. The computation of this matrix exponential therefore completes the calculation of both the QFI and its gradient.

\end{document}